\RequirePackage{rotating}
\documentclass[usenatbib]{mnras}
\usepackage{epsfig}
\usepackage{amsmath}
\usepackage{graphicx}
\usepackage{array}
\usepackage{textcomp}
\usepackage{amssymb}
\usepackage{rotating}
\usepackage{pdflscape}
\usepackage{caption}
\usepackage{subcaption}
\usepackage{longtable}
\usepackage{xtab}
\def\mnras {{\sc MNRAS}}
\def\apj {ApJ}
\def\apjs {ApJS}

\def\aj {AJ}

\def\aap {A\&A}

\title[Star Formation in Milky Way Analogues]
      {From the Outside Looking in: What can Milky Way Analogues Tell us About the Star Formation Rate of Our Own Galaxy?} 
\author[Fraser-McKelvie, Merrifield, \& Arag\'{o}n-Salamanca]
       {Amelia Fraser-McKelvie$^{1}$\thanks{Amelia.Fraser-McKelvie@nottingham.ac.uk}, Michael Merrifield$^{1}$, Alfonso Arag\'on-Salamanca$^{1}$
                \vspace*{1mm}\\
        $^{1}$ School of Physics \& Astronomy, University of Nottingham, University Park, Nottingham, NG7 2RD, U.K. \\
	}
\begin{document}
\maketitle
\begin{abstract}
\noindent{The Milky Way has been described as an anaemic spiral, but is its star formation rate (SFR) unusually low when compared to its peers?
To answer this question, we define a sample of Milky Way Analogues (MWAs) based on stringent cuts on the best literature estimates of non-transient structural features for the Milky Way. This selection yields only 176 galaxies from the whole of the SDSS DR7 spectroscopic sample which have morphological classifications in GZ2, from which we infer SFRs from two separate indicators. The mean SFRs found are $\log(\rm{SFR}_{SED}/\rm{M}_{\odot}~\rm{yr}^{-1})=0.53$ with a standard deviation of 0.23 dex from SED fits, and $\log(\rm{SFR}_{W4}/\rm{M}_{\odot}~\rm{yr}^{-1})=0.68$ with a standard deviation of 0.41 dex from a mid-infrared calibration. The most recent estimate for the Milky Way's star formation rate of $\log(\rm{SFR}_{MW}/\rm{M}_{\odot}~\rm{yr}^{-1})=0.22$ fits well within 2$\sigma$ of these values, where $\sigma$ is the standard deviation of each of the SFR indicator distributions. We infer that the Milky Way, while being a galaxy with a somewhat low SFR, is not unusual when compared to similar galaxies. } 
\end{abstract}
\begin{keywords}
Galaxy: general -- galaxies: evolution -- galaxies: general   -- galaxies: spiral -- galaxies: star formation
\end{keywords}
\section{Introduction}

Our privileged position within the Milky Way makes it difficult to determine basic physical properties which we can measure for external galaxies with relative ease. 
This makes it troublesome to establish whether the Milky Way is a peculiar galaxy, or extremely ordinary when placed on extragalactic scaling relations. Given the difficulty of observing the Milky Way from within, it is natural to turn to extragalactic analogues and study them to infer information about our own galaxy. 
The difficulty arises when we must define what an analogue galaxy is. Put another way, what are the defining characteristics of the Milky Way? 

Despite our embedded location, the main structural parameters of the Milky Way have been determined with some confidence \citep[see][for a thorough review]{Bland-Hawthorn16}. The spiral arm structure of the Milky Way's disk was first revealed by ionised Hydrogen distributions and observations of the 21cm line in the 1950s \citep[See][for a review of early work]{Oort58}. The Galactic bar took longer to elucidate due to the high extinction in the central regions of the Galaxy. Strong evidence only became available with the advent of large-scale photometric and spectroscopic surveys of the Galactic centre employing near-infrared (IR) photometry \citep{Hammersley94, Weiland94}, and gas kinematics \citep{Binney91}. Strong evidence for its existence was provided by the \textit{Spitzer Space Telescope} \citep{Werner04} in the 2000s \citep{Benjamin05} and several surveys since \citep[e.g.][]{Cabrera-Lavers08, Wegg15}. A small, boxy/peanut-shaped bulge was reported by \citet{Dwek95}, and numerous studies since have described bulge masses consistent with a low bulge-to-total ratio \citep[e.g.][]{Malhotra96, Binney97, Widrow08}, which is consistent with that found for nearby Sc-type galaxies \citep{Laurikainen07}.

It is probably true that the way a Milky Way analogue (MWA) sample is defined depends strongly on the particular property of the Milky Way of interest, and the science goals. Some examples include the colour \citep{Mutch11}, star formation rate \citep{Licquia15}, number of bright satellites \citep{Liu11,Robotham12}, luminosity and environment \citep{Geha17}, or a combination of all of these. Simulations have more freedom to enforce constraints on parameters such as galactic rotation curves \citep[e.g.][]{Bozorgnia16}, galaxy shapes \citep{Calore15}, and accretion histories \citep{Bullock05}. 

A popular method of selecting MWA samples is based on their star formation rate (SFR), or similarly, colour. The SFR of the Milky Way has been the subject of much previous work \citep[e.g.][]{Smith78, Misiriotis06, Davies11}, and in this work, we employ the estimate of \citet{Licquia15} who place the SFR of the Milky Way in the range $1.65\pm0.19~\textrm{M}_{\odot}~\textrm{yr}^{-1}$. This value, along with complementary colour estimates, lead the Milky Way to be considered to be undergoing a transition onto the red sequence \citep{Mutch11}.
The SFR of a galaxy may vary on short timescales \citep[e.g.][]{Davies15}, and hence selecting a sample of analogues to the Milky Way on this criterion alone will likely produce samples with heterogeneous structural parameters and internal physics. 

With this in mind, we set out to determine whether our Galaxy is indeed anaemic when compared to analogues selected in a structural manner, remaining agnostic to any transient observables such as SFR and colour. This analogue sample is selected from tight constraints on Milky Way structural parameters based on well-defined literature values, rendering a sample of MWAs as physically close as possible to what we expect for the Milky Way. To establish the viability of this approach, we seek to determine just how many analogues to the Milky Way exist within the largest spectroscopic sample of galaxies in the local Universe -- the Sloan Digital Sky Survey (SDSS). We calculate the SFR of these tightly-constrained structural analogues to determine whether the Milky Way is truly a galaxy outside the norm, or whether its star formation characteristics are simply to be expected for a galaxy of its type. 
 
In this paper, we convert all archival stellar mass and SFR estimates to a \citet{Kroupa02} IMF, and the standard $\Lambda$CDM cosmology is adopted with $H_{0} = 70 \rm{km}~\rm{s}^{-1}~\rm{Mpc}^{-1}$, $h=H_{0}/100$, $\Omega_{M}=0.3$ and $\Omega_{\Lambda}=0.7$

\begin{figure*}
\centering
\begin{subfigure}{0.99\textwidth}
\includegraphics[trim={0cm 1cm 3cm 1cm}, clip]{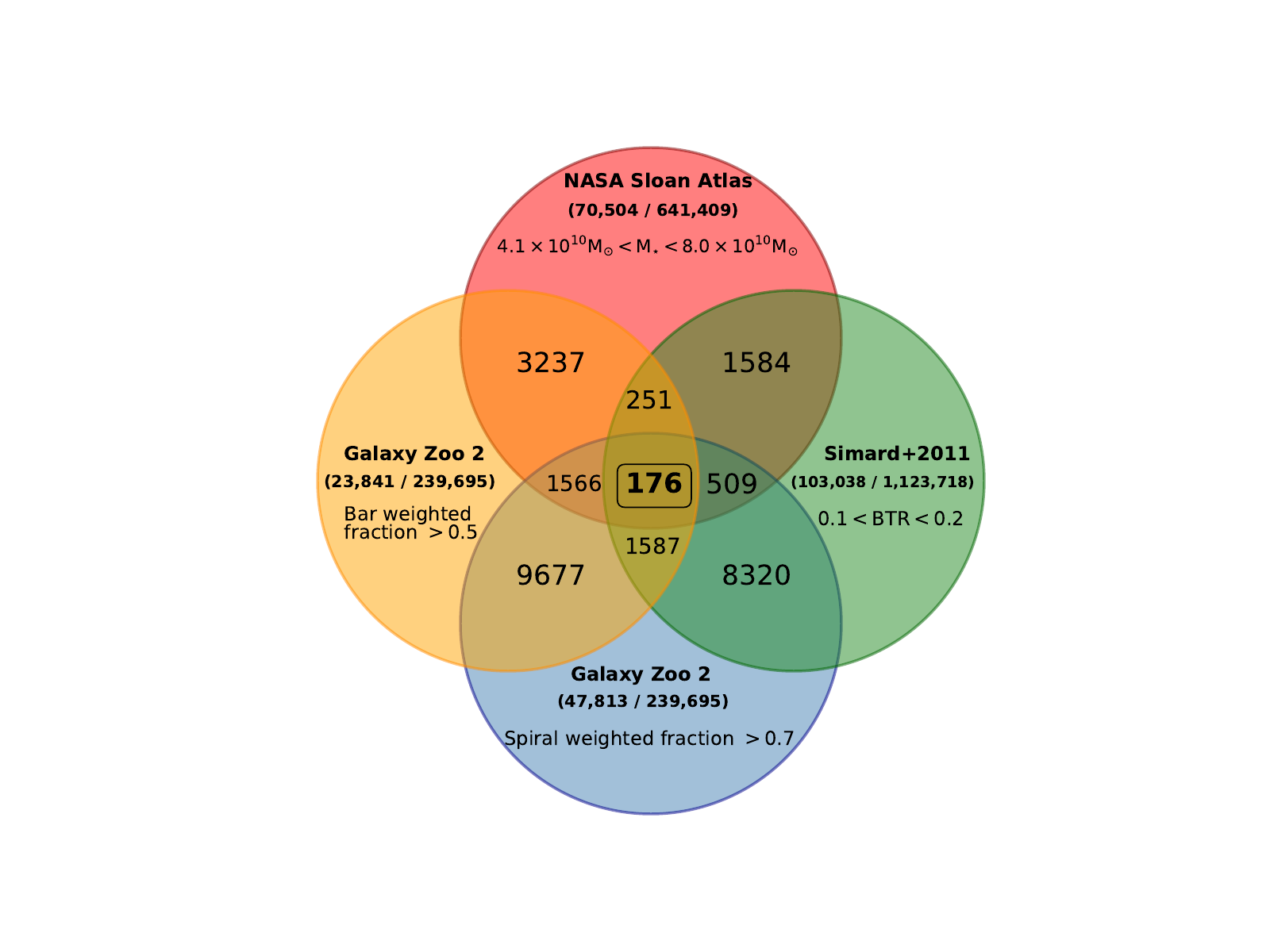} 
\end{subfigure}
\caption{A Venn diagram illustrating the overlap between the tight constraints imposed on the Milky Way Analogue sample and the number of galaxies from the NASA Sloan Atlas, Galaxy Zoo 2, and the \citet{Simard11} bulge-disk decomposition catalogues that satisfy each. The number of galaxies in each catalogue surviving the cut applied, along with the original number of galaxies in each catalogue is shown in brackets. Just 176 galaxies satisfy all four criteria.} 
\label{Venn}
\end{figure*}

\begin{figure}
\centering
\begin{subfigure}{0.99\textwidth}
\includegraphics[width=0.49\textwidth]{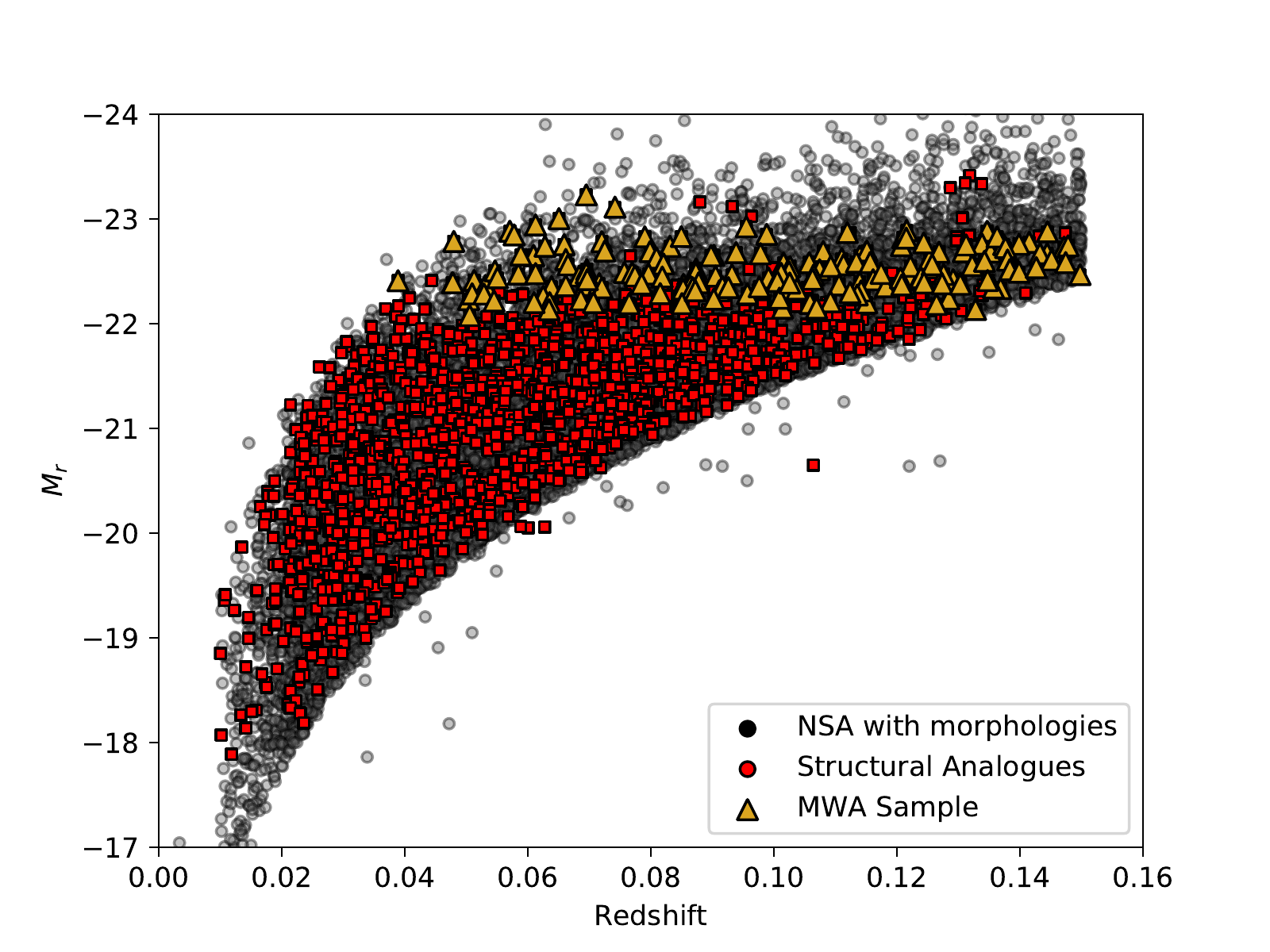} 
\end{subfigure}
\caption{NSA $r$-band absolute magnitude as a function of redshift for the NSA catalogue with associated GZ2 morphologies (black points), and structural Milky Way analogues with the same BTR, spiral arm and bar cuts used for the MWA sample selection (red squares). The structural analogues are a representative sample of the overall NSA catalogue. The MWA sample (with the additional mass range criterion) selected by the structural criteria listed in Section~\ref{SS} are shown as yellow triangles. Thanks to the brightness of the Milky Way, the MWA sample is essentially complete out to $z=0.15$.} 
\label{completeness}
\end{figure}

\begin{figure*}
\centering
\begin{subfigure}{0.27\textwidth}
\includegraphics[width=\textwidth]{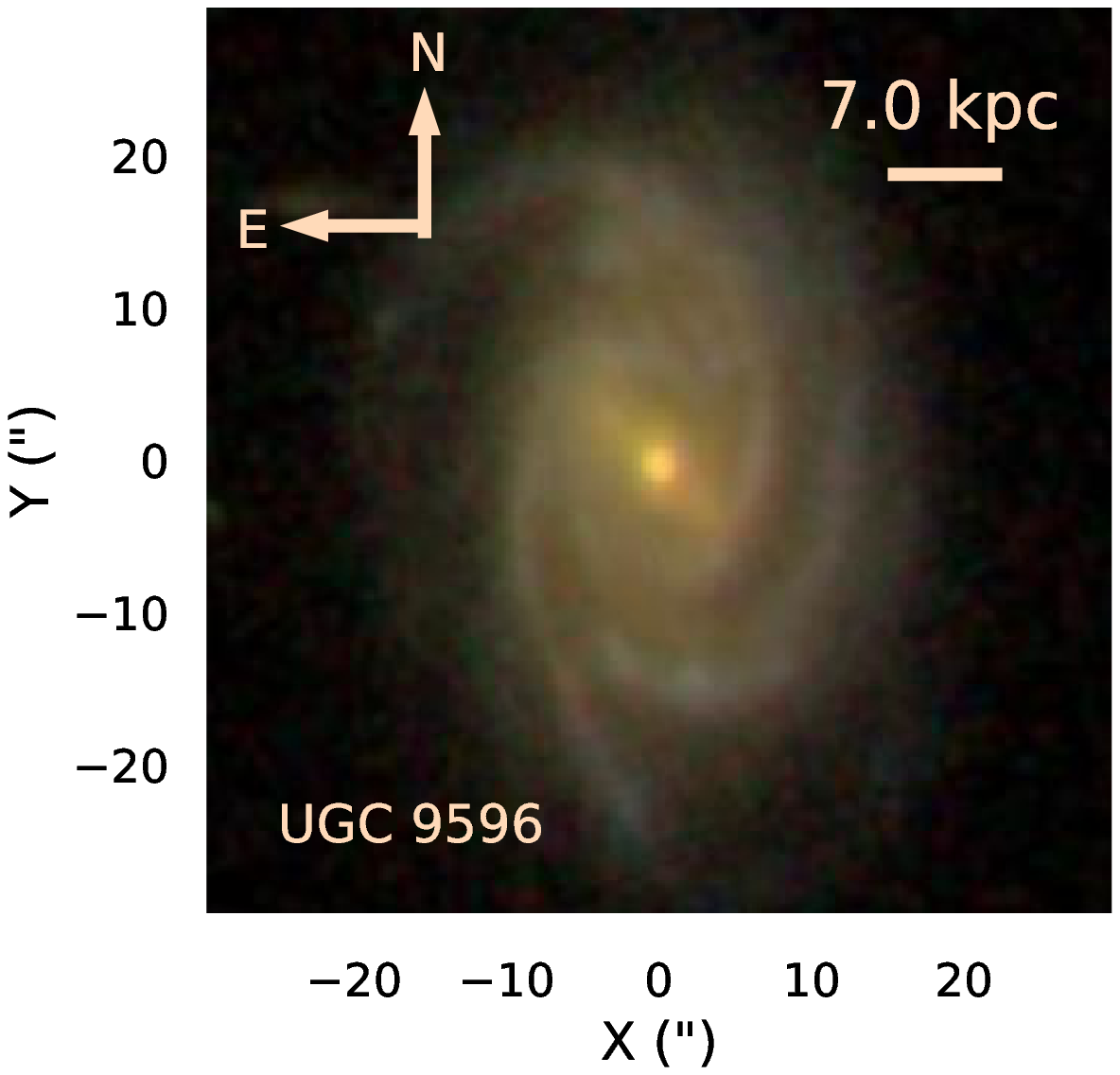} 
\end{subfigure}
\begin{subfigure}{0.27\textwidth}
\includegraphics[width=\textwidth]{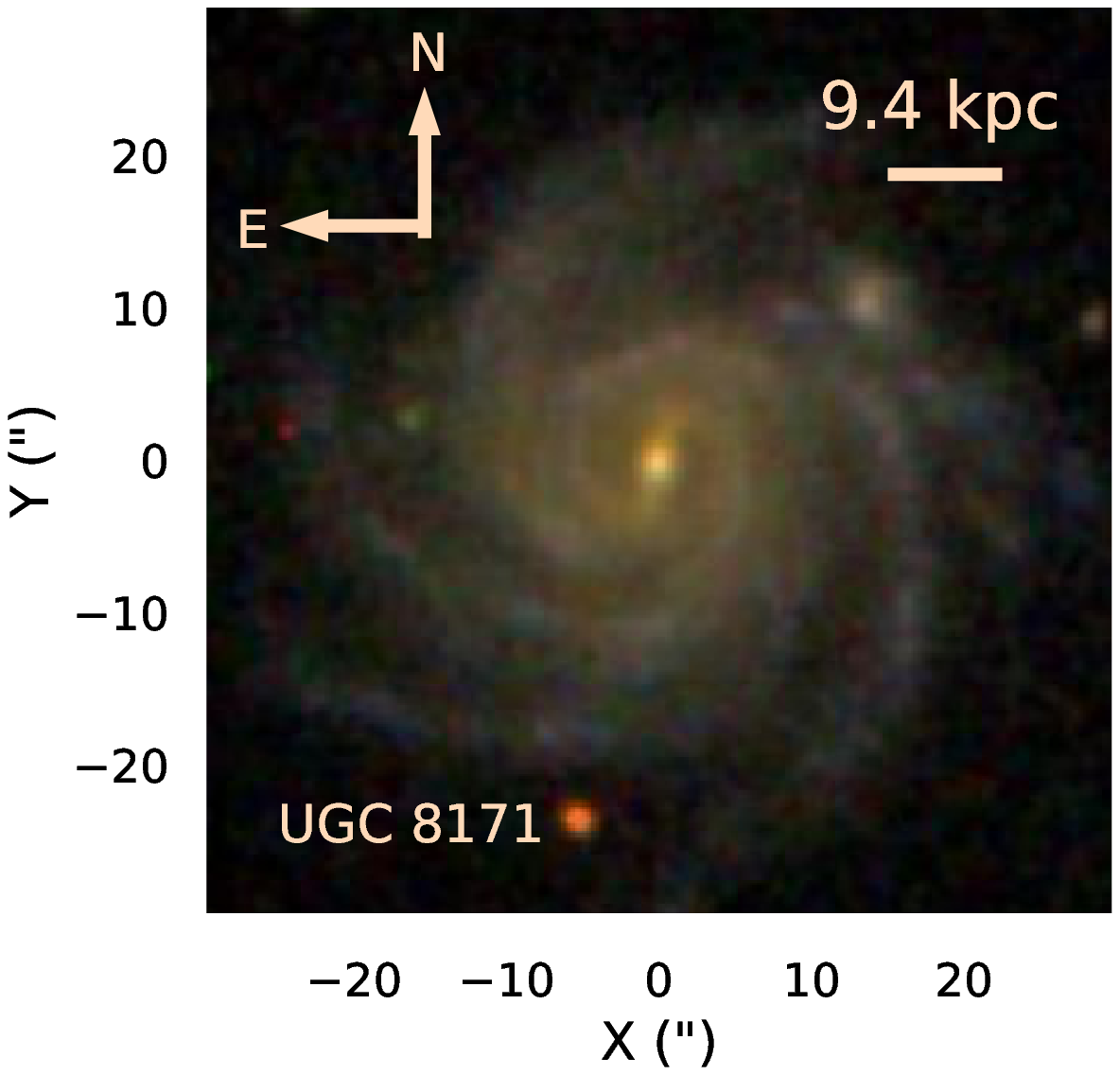} %
\end{subfigure}
\begin{subfigure}{0.27\textwidth}
\includegraphics[width=\textwidth]{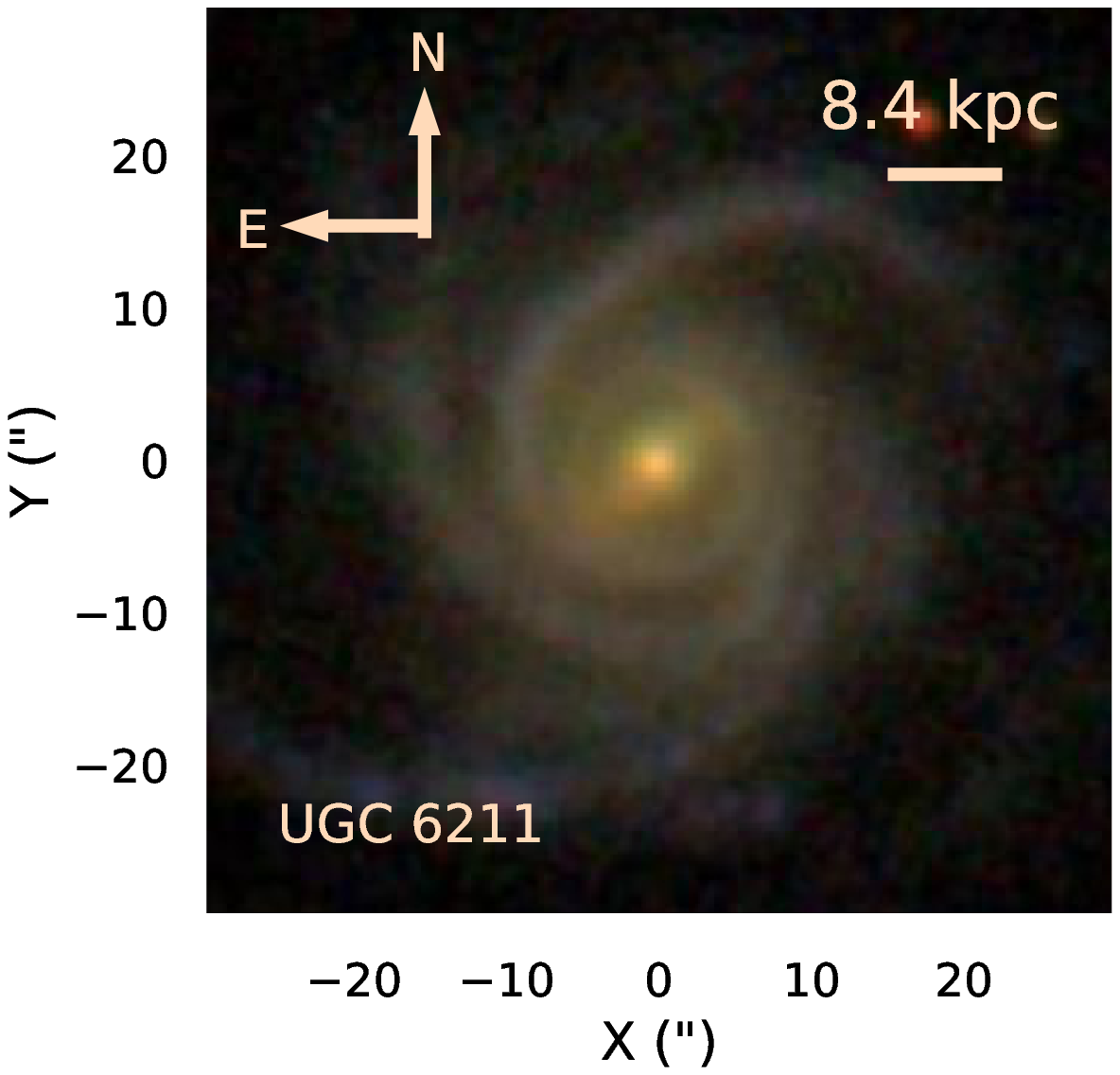} %
\end{subfigure}
\hfill
\begin{subfigure}{0.27\textwidth}
\includegraphics[width=\textwidth]{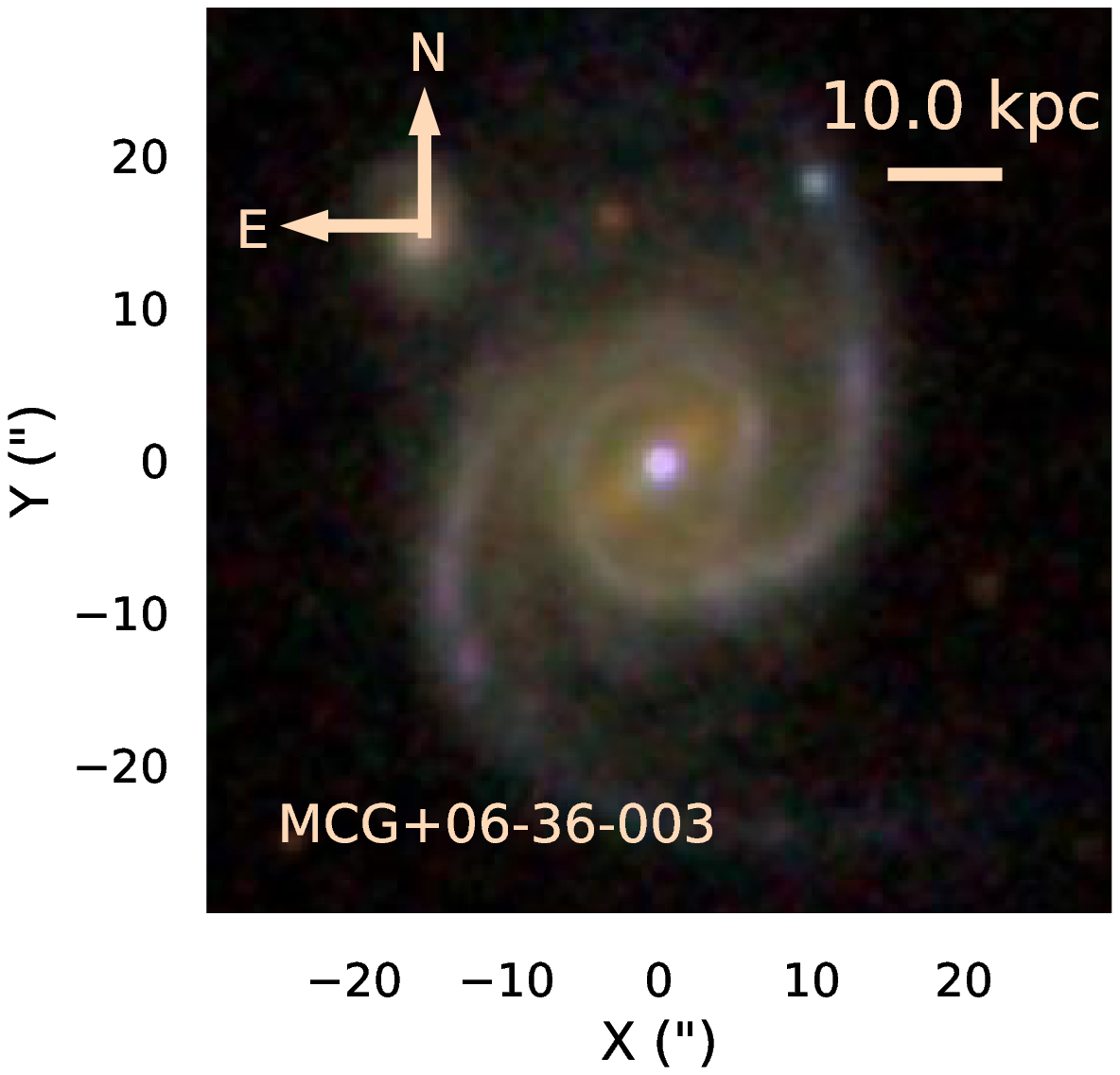} %
\end{subfigure}
\begin{subfigure}{0.27\textwidth}
\includegraphics[width=\textwidth]{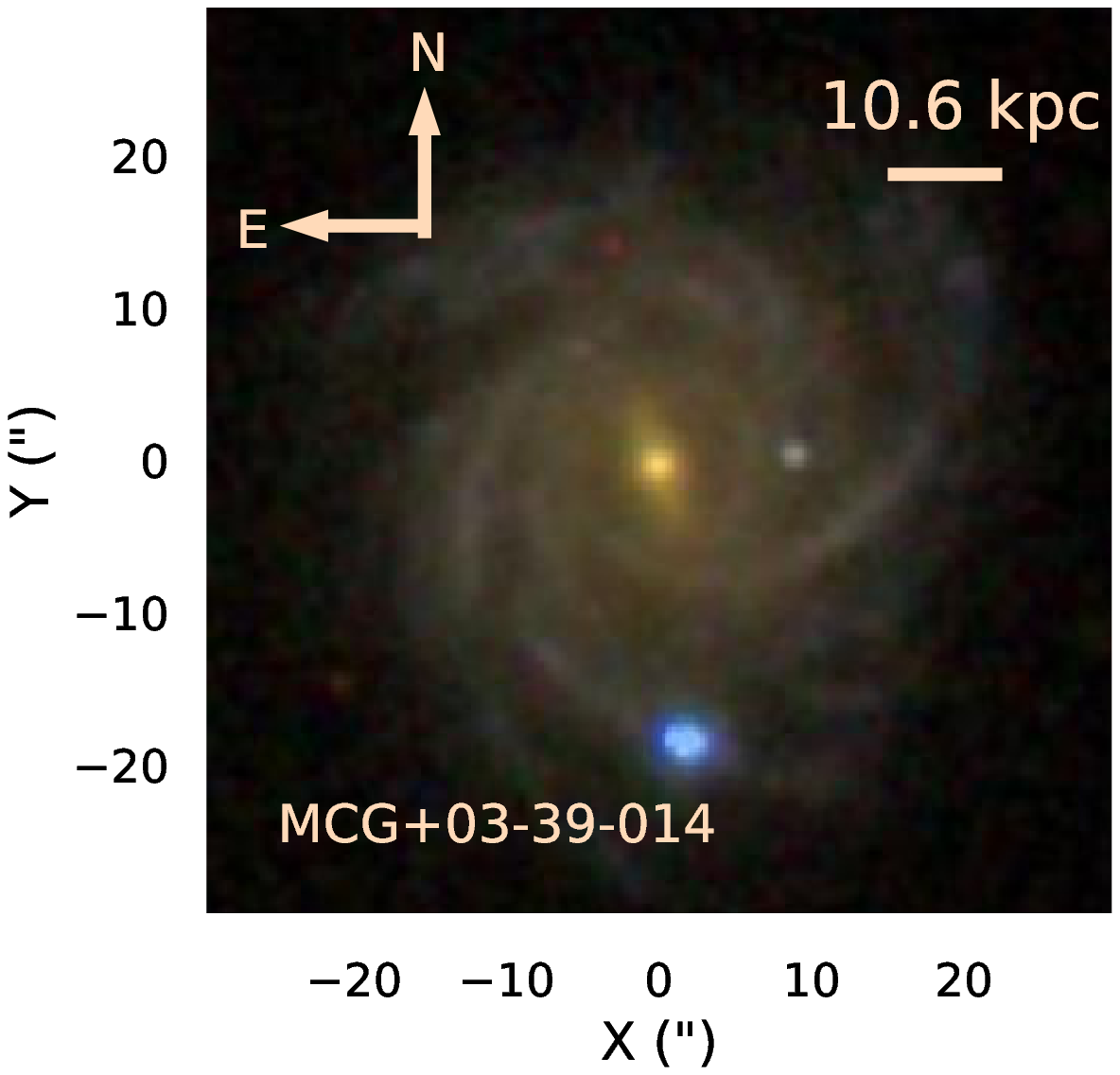} %
\end{subfigure}
\begin{subfigure}{0.27\textwidth}
\includegraphics[width=\textwidth]{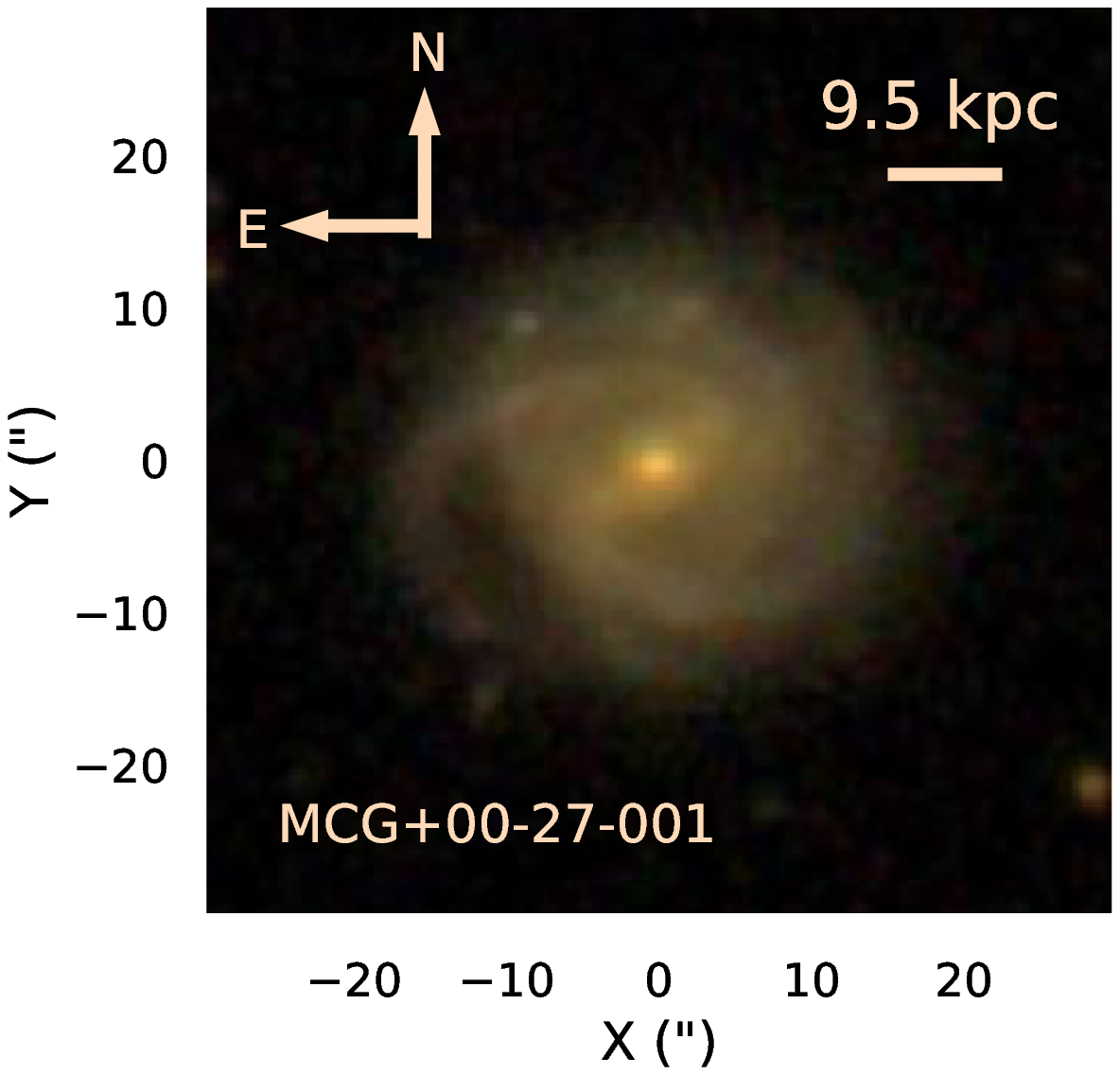} %
\end{subfigure}
\hfill
\begin{subfigure}{0.27\textwidth}
\includegraphics[width=\textwidth]{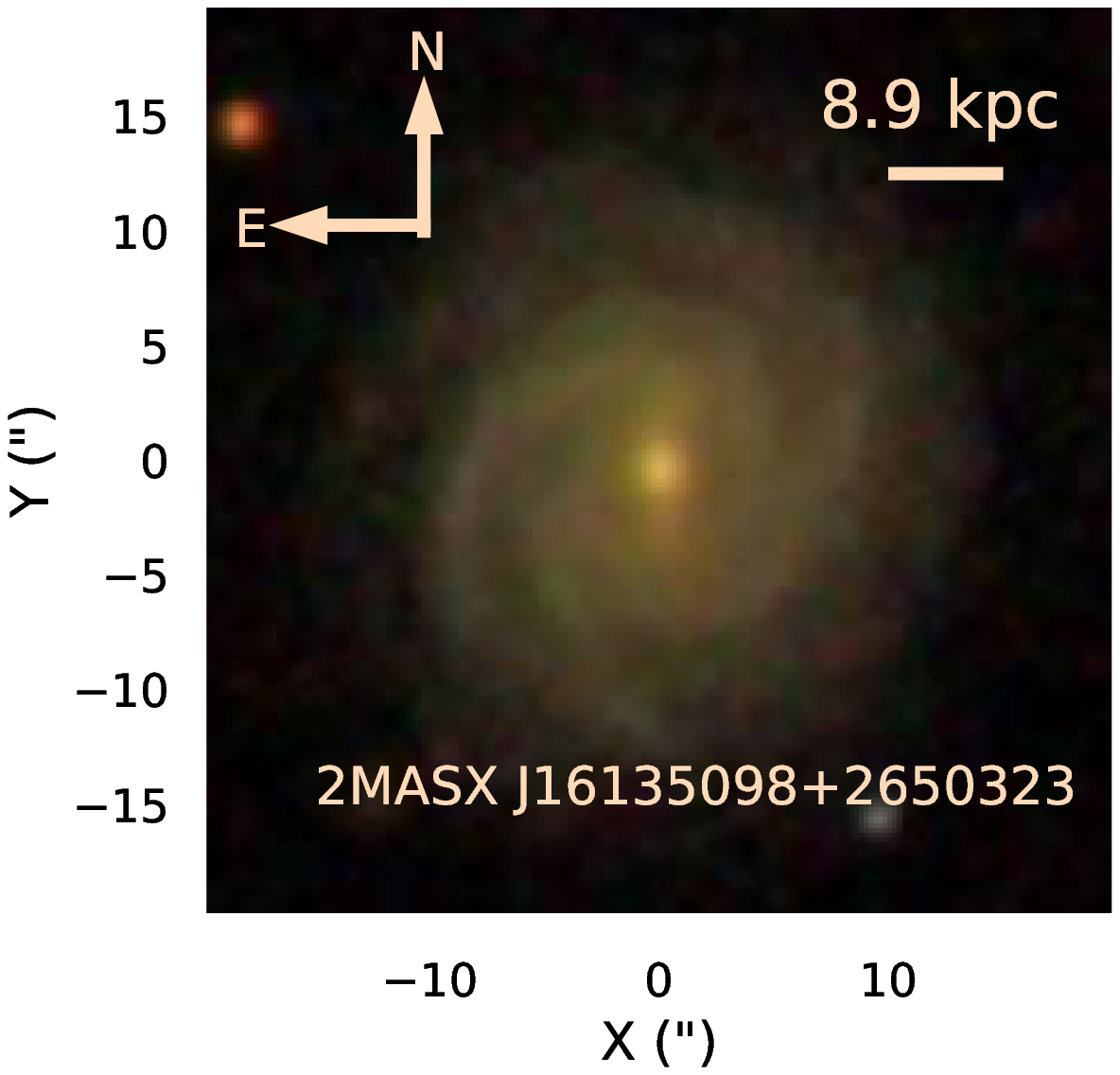} %
\end{subfigure}
\begin{subfigure}{0.27\textwidth}
\includegraphics[width=\textwidth]{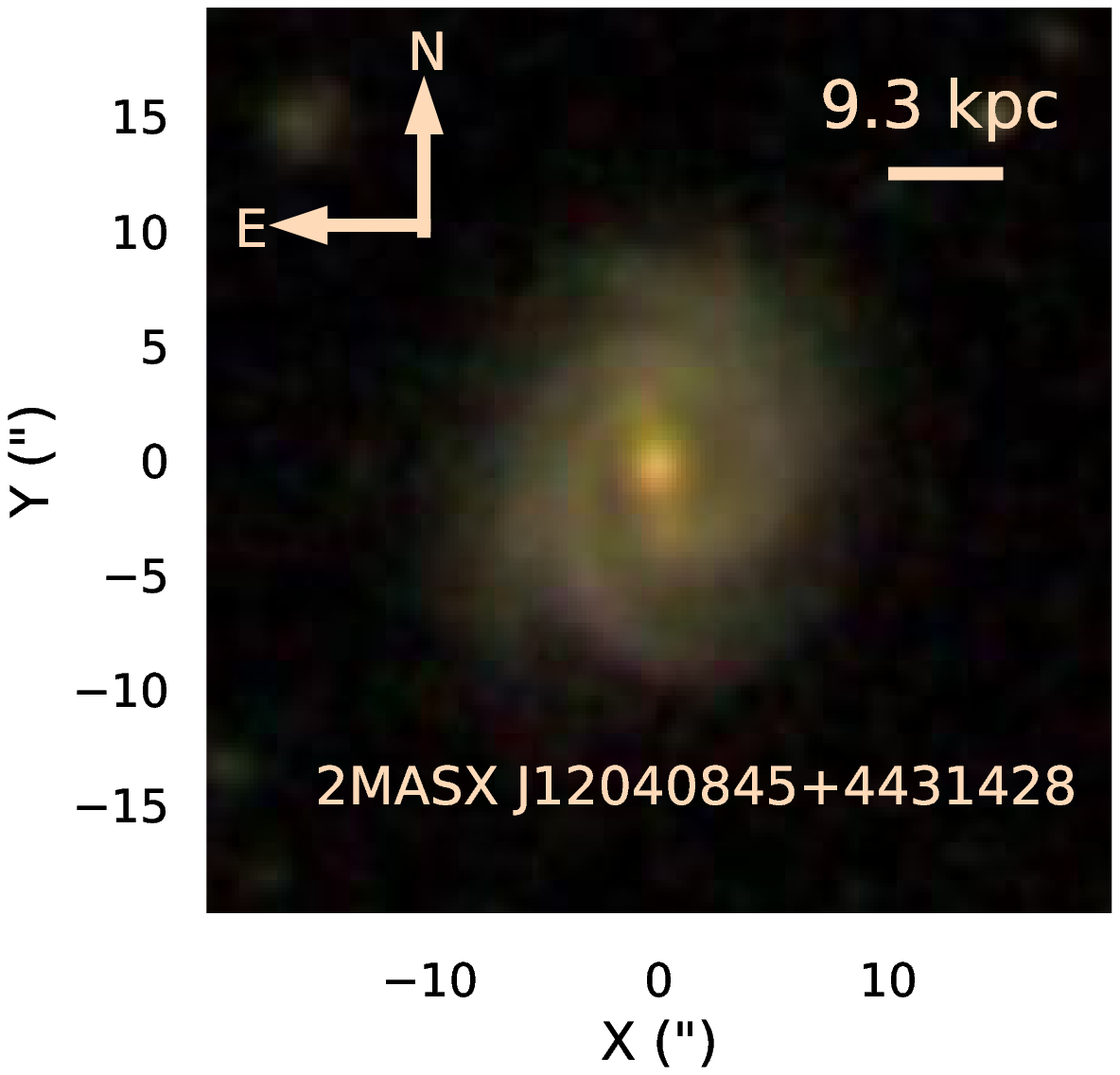} %
\end{subfigure}
\begin{subfigure}{0.27\textwidth}
\includegraphics[width=\textwidth]{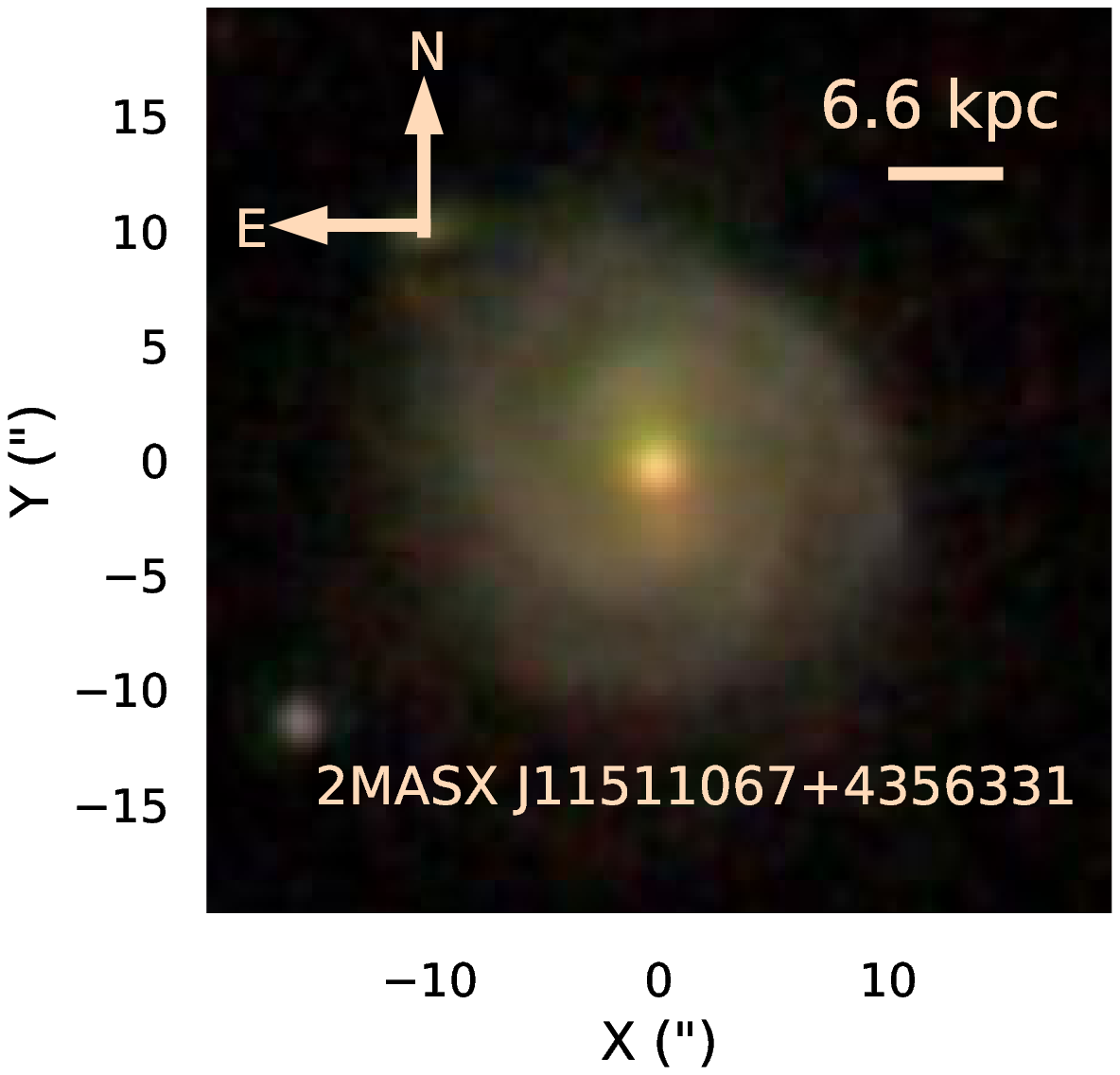} %
\end{subfigure}
\hfill

\caption{SDSS $gri$ colour images of nine Milky Way analogue examples, chosen to be between a narrow mass range, possess a small bulge, spiral arms, and a bar.}
\label{pretty_pics}
\end{figure*}

\section{Milky Way Analogue Sample Selection}
\label{SS}
To create a sample of Milky Way analogues, we team a tight constraint on stellar mass with three structural parameters that are well-determined for the Milky Way: the presence of spiral arms, a bar, and a small bulge. Only a galaxy satisfying all four criteria will be termed a MWA. The parent sample is the largest spectroscopic redshift survey with an abundance of ancillary data products and value-added catalogues: the Sloan Digital Sky Survey Data Release 7 \citep[SDSS DR7;][]{Abazajian09}, a redshift survey that obtained photometry and spectroscopy for over a million galaxies. This parent sample was chosen not only for the abundance of supplementary data, but the sheer number statistics which allow us to make extremely stringent cuts. The parameter cuts and relevant catalogues derived from SDSS DR7 galaxies are:
\begin{itemize}

\item \textbf{The presence of spiral arms}. For the spiral and bar classifications, we utilise the Galaxy Zoo 2 \citep[GZ2;][]{Willett13} citizen science project, which obtained classifications for bright and nearby galaxies from SDSS images. The catalogue of \citet{Hart16} was employed, which contains both original and redshift-debiased estimates of structural parameters for 239,695 galaxies with m$_{r}\eqslantless17$ with spectroscopic redshifts from SDSS DR7. We utilise the spiral fraction parameter and choose a cut such that 70\% of respondents classified a particular galaxy as having spiral arms, weighted by the accuracy of a given respondent. This spiral cut was determined after visually examining samples with both higher and lower vote fraction counts. This value was chosen as it provided the greatest number of spiral galaxies with a small amount of contamination to the sample. The GZ2 parameter was \texttt{t04\_spiral\_a08\_spiral\_weighted\_fraction $>$ 0.7}. \\

\item \textbf{The presence of a bar}. 
Again, we used GZ2 structural parameters to determine the presence of a bar. We used \texttt{t03\_bar\_a06\_bar\_weighted\_fraction $>$ 0.5}, as in \citet{Masters12}. The combination of this bar threshold and the spiral arm fraction threshold were most likely to produce barred spiral galaxies. We note that there are likely plenty of barred galaxies and spiral galaxies below the thresholds chosen, but there are also edge-on galaxies masquerading as bars, and ringed or irregular galaxies that are classified as spiral. To ensure an uncontaminated sample, these thresholds were chosen. 
This may also have the effect of biasing the sample towards strongly barred galaxies, which may be more quiescent than the average galaxy population \citep{Masters11, Fraser-McKelvie18}, although, \citet{Masters12} shows that 90\% of strong and intermediate bars classified by \citet{Nair10} are recovered using the threshold chosen. We interpret this to mean that GZ2 bar classifications do no worse than human classification of optical images. \\

\item \textbf{Bulge-to-Total Ratio (BTR):} \citet{Licquia15}, report the Milky Way's BTR to be $0.131 < \rm{BTR} < 0.178 $, which is a stellar mass ratio based on a hierarchical Bayesian statistical analysis and given tight priors on Milky Way bulge and disk masses from literature. 
We adopt the BTRs from the \citet{Simard11} catalogue of photometric bulge+disk decompositions for SDSS galaxies. We use the $r$-band photometric decompositions, with the S\'{e}rsic index of the bulge, $n_{b}$, set as a free parameter (which gives better decompositions for small bulges than the alternative of setting $n_{b}=4$). This catalogue is for all galaxies in the Legacy survey of the SDSS DR7 \citep{York00}, regardless of redshift, and contains 1,123,718 galaxies. The average error in BTR for BTRs in the range of the Milky Way is $\sim0.02$, and taking this into account by adding this error to the uncertainty of the \citet{Licquia15} estimate in quadrature, the 1$\sigma$ range becomes $0.1<\rm{BTR}<0.2$. We note that while the value for the Milky Way is a stelar mass ratio, the \citet{Simard11} BTRs are based on $r$-band photometry and will hence be luminosity ratios. These however should be comparable to the stellar mass ratio of \citet{Licquia15}, as for galaxies of a similar mass the mass to light ratio should not change by a large amount. \\

\item \textbf{Stellar Mass:} \citet{Licquia15} determine the mass of the Milky Way to be in the range $4.94\times10^{10}~\textrm{M}_{\odot}<\textrm{M}_{\star}<7.22\times10^{10}~\textrm{M}_{\odot}$, with a Kroupa IMF. We note this mass range also wholly encompasses the mass range for the Milky Way presented in the Bayesian analysis of \citet{McMillan11}. 

To obtain stellar mass estimates of SDSS galaxies, we employ the NASA-Sloan Atlas (NSA) catalogue \citep{Blanton11} masses derived from elliptical Petrosian photometry, recommended for extended sources such as nearby galaxies with a Chabrier IMF. Stellar masses in this catalogue are generated using \textsc{kcorrect} \citep{Blanton07}, which provides an estimate of the current stellar mass of a galaxy from a library of template SEDs generated by SSP models. The advantage of \textsc{kcorrect} is it provides a nearly model-independent estimate of stellar mass. 
The NSA is a reanalysis of SDSS photometry that incorporates better sky subtraction and deblending, which particularly aids in the analysis of larger galaxies. The elliptical Petrosian photometry, along with an increase in redshift range, was added originally for the targeting catalogue of the Mapping Nearby Galaxies at Apache Point Observatory (MaNGA) galaxy survey. SDSS Data Release 13 contains the new version of the NSA, \texttt{v1\_0\_1}, which consists of 641,409 bright, nearby galaxies. 

While in theory, the elliptical Petrosian magnitudes should recover essentially all of the flux of an exponential galaxy profile \citep{Stoughton02}, it is important to note that all studies of the Milky Way and comparisons to external galaxies will suffer some level of unknown bias resulting from differing methods of measurement of physical properties (for example stellar masses, BTRs and SFRs). It is important to keep in mind the caveat that it is difficult to compare measurements based on individual stars and resolved gas to unresolved extragalactic measurements.

The public NSA catalogue does not include any uncertainties on the stellar mass estimates, so assuming the photometric error is negligible compared to the uncertainty in the stellar mass-to-light ratio (M/L), we estimate the error on stellar mass from the range of M/L for a galaxy with colour of a typical spiral from different exponentially-declining star formation rate models of age 12 Gyr from \citet{Bell01}. The spread in M/L between the models listed and for a given IMF is $\sim0.1 \rm~{dex}$, corresponding to a systematic error in stellar mass determination of 25\%. We add this in quadrature to the uncertainty in the \citet{Licquia15} measurement to obtain the 1$\sigma$ range in stellar mass about the Milky Way value. The stellar mass range used for this work is therefore $4.1\times10^{10}~\textrm{M}_{\odot}<\textrm{M}_{\star}<8.0\times10^{10}~\textrm{M}_{\odot}$.
\\ 
\end{itemize}

A Venn diagram in Figure~\ref{Venn} shows the numbers of MWAs for each combination of cuts. The most restrictive parent catalogue was GZ2, as its original sample selection was only the brightest $\sim$25\% of resolved galaxies in SDSS DR7 \citep{Willett13}, leaving just 239,695 galaxies. From Figure~\ref{Venn}, the most restrictive cut is the mass cut. The resultant sample is low redshift, limited to the range $0.04 <z< 0.15$, the upper limit being set by the redshift limit of the NSA, and no bounds on the lower limit. 

We investigate the completeness of the MWA sample in Figure~\ref{completeness}. Here we plot the $r$-band absolute magnitude against redshift for NSA galaxies with morphological classifications from GZ2 ($\sim$215,000 galaxies, black points), all galaxies that satisfy the MWA structural selection criteria of the BTR range and GZ2 spiral arm and bar fraction limits (1587 galaxies, red squares) and the final MWA sample with a mass range cut (176 galaxies, gold triangles). We see that the sample is complete within all of the NSA with associated GZ2 morphologies, and the structural parameters imposed.
The structural analogues are a representative sample of all of the morphology-matched NSA catalogue, and when an additional mass cut is added, we see the sample of MWAs is essentially complete out to $z=0.15$. By incorporating the extremely strict structural criteria listed above, we end up with a well-defined sample of 176 galaxies in SDSS DR7, a selection of which are shown in Figure~\ref{pretty_pics}.

\section{Star Formation Rates}
\label{SFR_I}
\subsection{Star Formation Rate of the Milky Way}
The SFR range for the Milky Way of \citet{Licquia15} of $1.65\pm0.19~\textrm{M}_{\odot}~\textrm{yr}^{-1}$ is used in this work. This range is drawn from the thorough review of literature values of \citet{Chomiuk11} who take archival literature values of the Milky Way SFR from a combination of physical indicators including infrared diffuse emission and point sources, massive star counts and supernova rates, and Lyman continuum photon rates. \citet{Chomiuk11} attempt to homogenise this sample by correcting each literature value to a common IMF (Kroupa) and stellar population synthesis model (\textsc{Starburst99}). 

\citet{Licquia15} take the SFR values and associated uncertainties from \citet{Chomiuk11} and statistically combine them and their associated uncertainties using hierarchical Bayesian modelling to produce a posterior distribution of the star formation rate. 
In this way, tension in the uncertainties associated with literature values may be accounted for, and outlier values from incorrect measurements or those that suffer from unaccounted for systematic errors will affect the resultant PDF less than for simpler statistical techniques, for example, the mean.

\subsection{Milky Way Analogue Star Formation Rate Indicators}
We take a multi-wavelength approach to calculating SFRs of the MWAs, using both UV/Optical/mid-IR spectral energy distributions (SEDs), and as a cross-check, a mid-infrared (IR) indicator calibrated to total-IR luminosity. Given these SFR indicators employ contrasting techniques, and overlapping but not identical wavelength ranges, they are complementary to one another.

UV/Optical/mid-IR SFRs are provided by the \textit{GALEX}-SDSS-\textit{WISE} Legacy Catalogue 2 \citep[GSWLC-2;][]{Salim16, Salim18}, which provides SFRs for 659,229 galaxies within SDSS with $z<0.3$. We utilise the GSWLC-X2 catalogue, which uses the deepest \textit{GALEX} photometry available (selected from the shallow `all-sky', medium-deep, and deep catalogues) for a source in the SED fit. SED fitting was performed using the Code Investigating GALaxy Emission \citep[CIGALE;][]{Noll09,Boquien19}, which constrains SED fits with IR luminosity, which they term SED+LIR fitting. The advantage of an SED fit over the traditional H$\alpha$ flux-derived SFRs is that the whole shape of the spectrum is used to derive SFRs, instead of the often dust-obscured monochromatic Balmer lines. SED templates will also account for any emission due to active galactic nuclei, ensuring that star formation rate estimates will include only gas ionized by young stars. 149 of the 176 galaxies in the MWA sample have matches in the GSWLC-X2 catalogue.

To act as a comparison, we also employ two mid-IR SFR calibrations. The mid-IR detects reprocessed emission from dust heated by young stars, and therefore is insensitive to the effects of dust attenuation. The \textit{Wide-field Infrared Survey Explorer} \citep[\textit{WISE};][]{Wright10} surveyed the sky in four infrared bands: 3.4, 4.6, 12, and 23$\mu$m, denoted bands W1-W4. The W3 band detects emission primarily from polycyclic aromatic hydrocarbons excited by UV radiation from young stars, though also traces the stellar continuum. The W4 band traces chiefly emission from dust heated by young stars, and \citet{Cluver14} demonstrated the reliability of the both the W3 and W4 bands to act as indirect indicators of star formation.

We use the most recent calibrations of \citet{Cluver17}, which includes a prescription to mitigate the effects of emission from old stellar populations at the tail of the Rayleigh-Jeans spectrum from the W3 and W4 bands. \citet{Cluver17} determine that 15.8\% of W3 band flux and 5.9\% of W4 band flux are contaminated by stellar emission. Using the W1 band as a proxy for stellar emission, in a similar method to \citet{Helou04}, they subtract these fractions of W1 flux from the W3 and W4 bands before calculating luminosities. The \citet{Cluver17} relation is calibrated with a number of bright, nearby galaxies with a range of luminosities, stellar masses and SFRs to total-IR luminosity measurements obtained from \textit{Spitzer} and \textit{Herschel} data. 

Photometry is taken from the AllWISE Source Catalogue \citep{Cutri14}, which contains photometry for over 747 million sky objects. Sources are matched to their SDSS MWA counterparts on the sky, and the correct aperture chosen based on the angular extent of the source as indicated by the extended source flag in the AllWISE catalogue. If the source was deemed `extended' on the sky (as in 77\% of cases), extended aperture photometry was employed, using apertures scaled to the 2MASS shape of the galaxy. If, however, the galaxy was a point source, the profile fit magnitudes were used.
In a small number of cases, $WISE$ photometry was not available for a galaxy, due primarily to the AllWISE photometric pipeline splitting extended sources into two separate objects, or scattered moonlight contamination. In these cases, the mid-IR SFR was not calculated for that galaxy. Redoing photometry by hand would alleviate this problem, and we note the further caveat that WISE catalogue photometry likely underestimates the true flux of an object. This is particularly the case for the extended aperture magnitudes. 157 and 155 of the 176 MWAs had W3 and W4 photometry available from the AllWISE catalogue, respectively. The majority of those missing were due to source splitting by the AllWISE pipeline. Five MWAs had neither GSWLC-X2 nor WISE-derived SFRs. 

\section{Results \& Discussion}
\label{Results}

\begin{figure*}
\centering
\begin{subfigure}{0.32\textwidth}
\includegraphics[width=\textwidth]{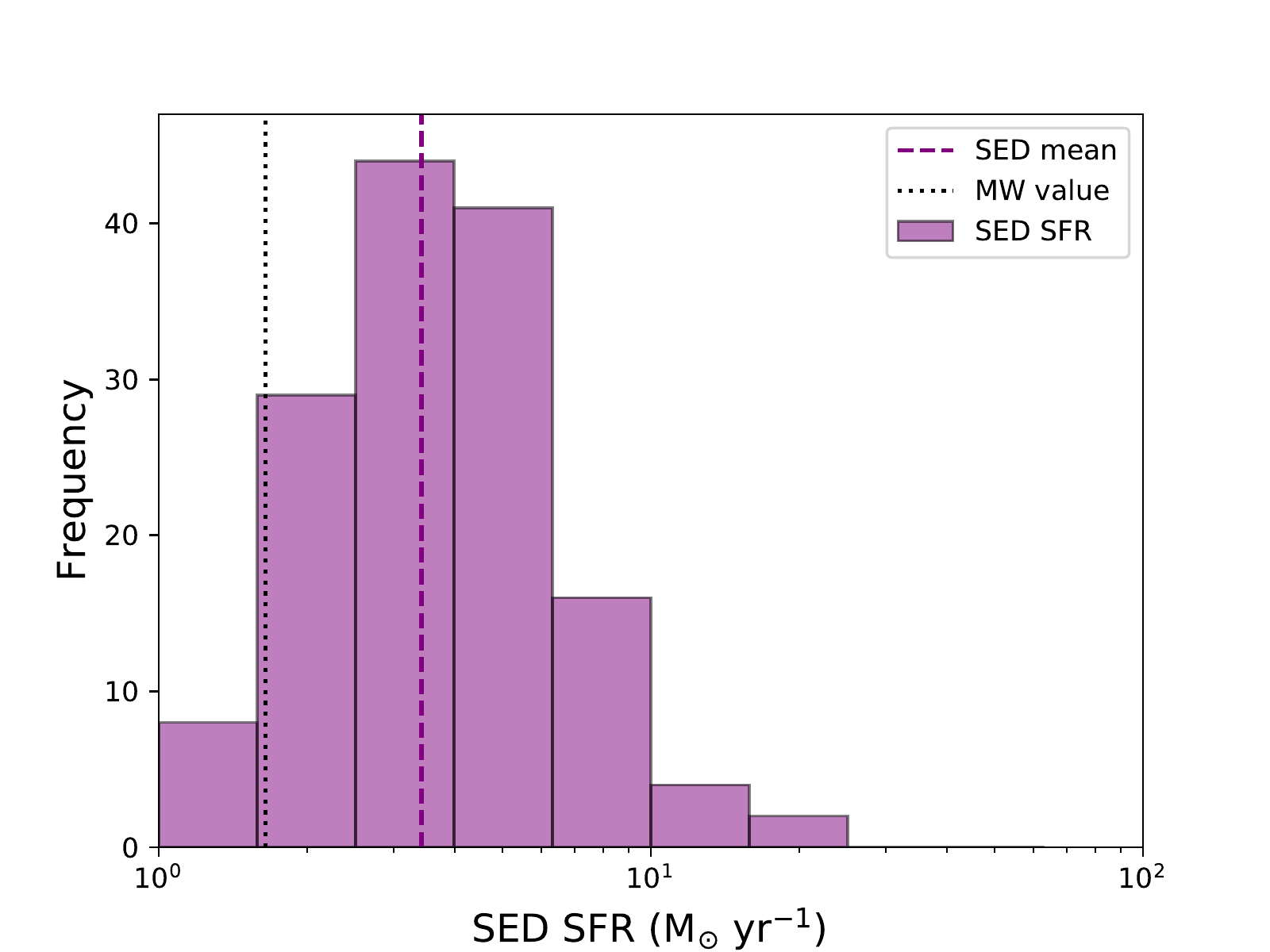} %
\caption{SED-derived SFRs.}
\end{subfigure}
\begin{subfigure}{0.32\textwidth}
\includegraphics[width=\textwidth]{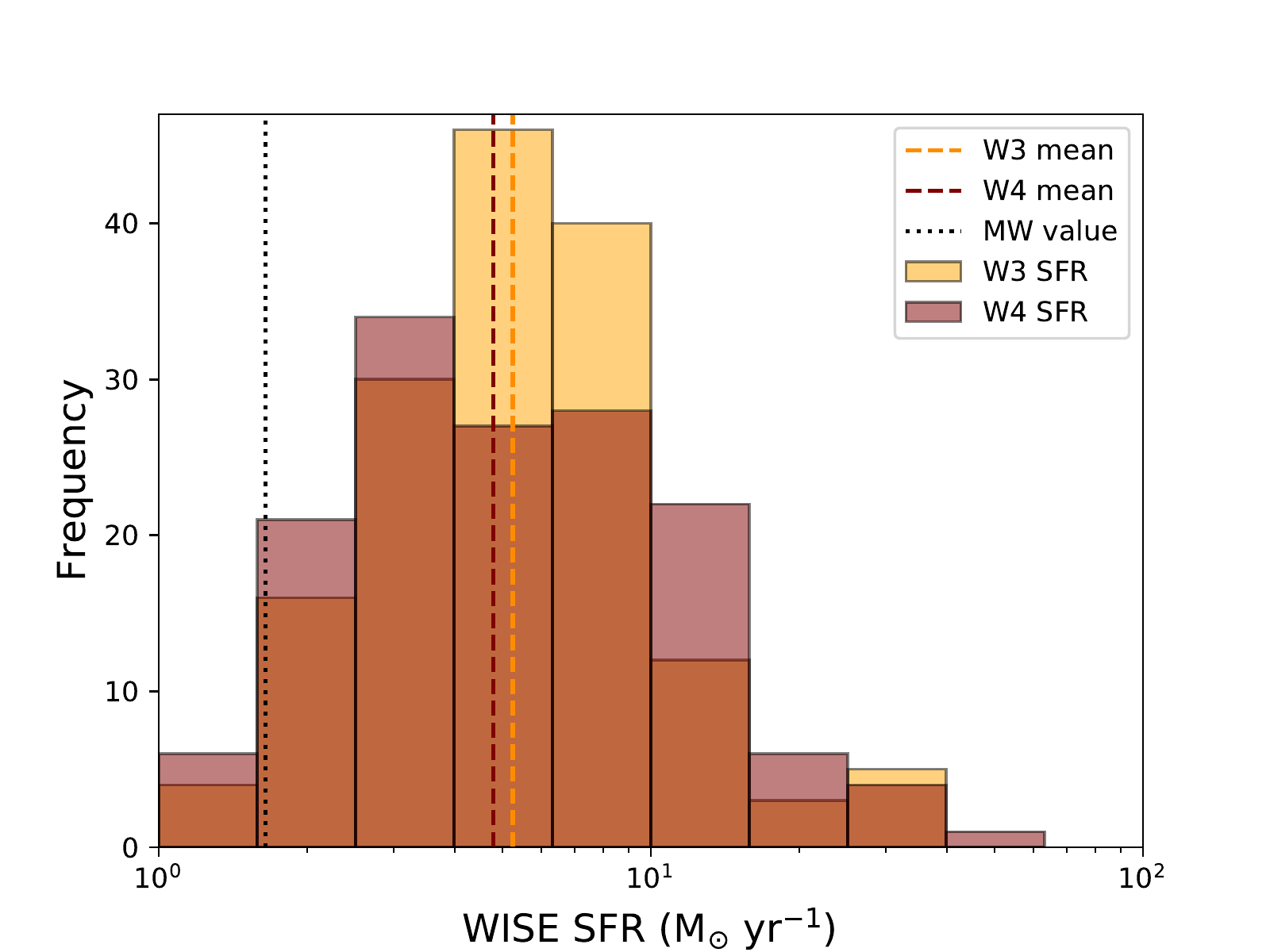} %
\caption{WISE W3 and W4 SFRs.}
\end{subfigure}
\begin{subfigure}{0.32\textwidth}
\includegraphics[width=\textwidth]{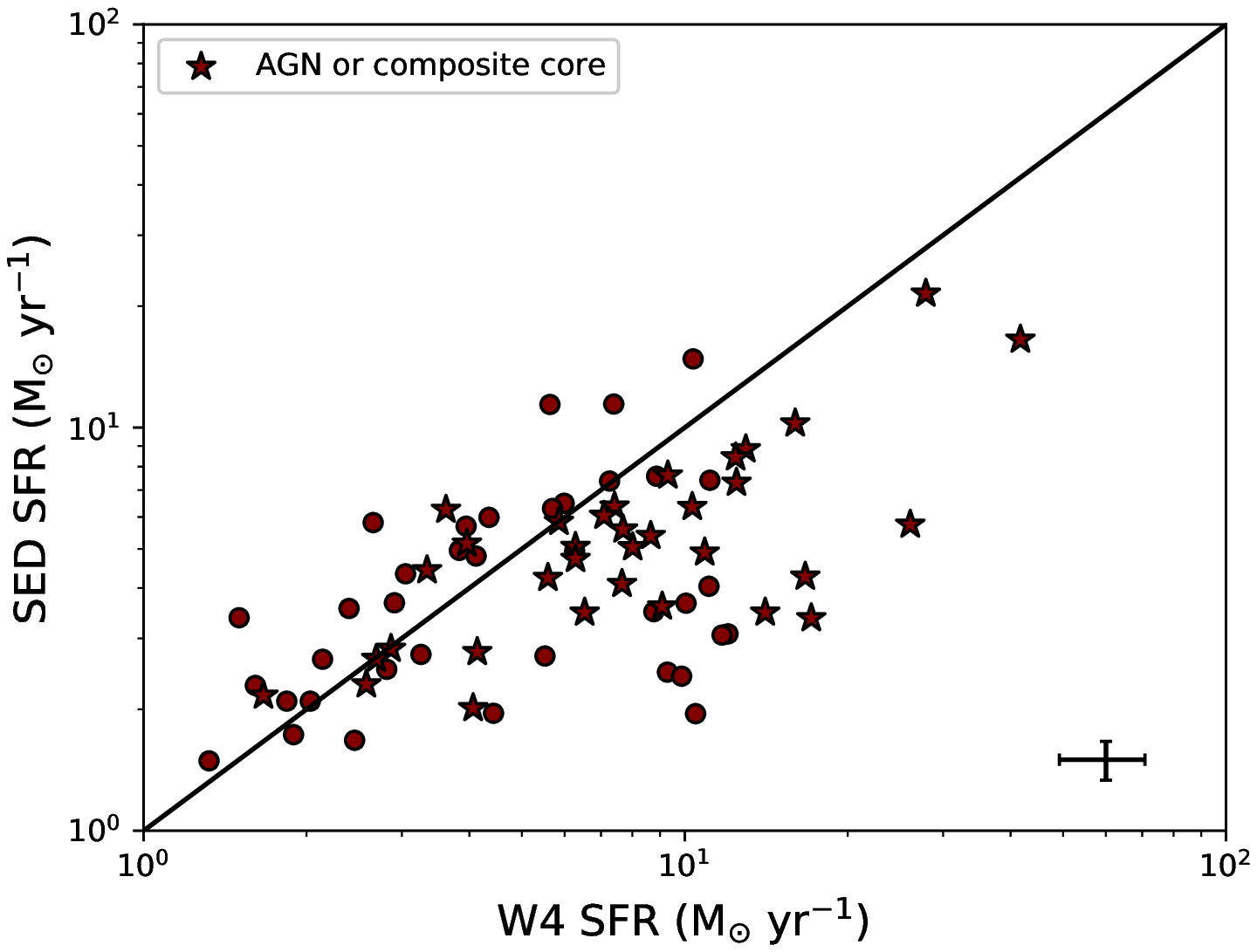} %
\caption{Comparison between W4 and SED-derived SFRs.}
\end{subfigure}
\hfill
\caption{Star formation rates of the Milky Way Analogue sample. Panel a) is a histogram of the SED-derived SFRs of \citet{Salim18} converted to a Kroupa IMF. The mean of this distribution is denoted by a dashed line and the Milky Way value from \citet{Licquia15} shown as a black line. For comparison, panel b) is a histogram of the \textit{WISE} W3- and W4-derived SFRs using the relation of \citet{Cluver17}, again, with the mean values denoted by dashed lines, and the Milky Way value shown in black. Panel c) directly compares the SFR measurements for the MWA sample galaxies in both the W4 and SED-derived SFR indicators. Optically-classified AGN or composite core regions from the catalogue of \citet{Brinchmann04} are shown as stars. There is significant scatter between the SED-derived and W4 SFR indicators, and the $WISE$-derived SFRs seem to be systematically offset higher than the SED-derived SFRs, possibly due to AGN contamination in the $WISE$ bands. Regardless, a general trend remains between these indicators.} 
\label{SFR_plots}
\end{figure*}

In Figure~\ref{SFR_plots}, we present the derived SFRs of the MWA sample, where the SED-derived SFRs have been converted from a Chabrier to a Kroupa IMF using the conversion as in \citet{Zahid12}. A histogram of the SED-derived SFRs of \citet{Salim18} is shown in panel a), with a mean of $\log(\rm{SFR}_{SED}/\rm{M}_{\odot}~\rm{yr}^{-1})=0.53$ and standard deviation of 0.23 dex, where these values are averages and standard deviations on the logarithmic SFRs. Panel b) shows the mid-IR-derived SFRs using the \textit{WISE} W3 and W4 calibrations of \citet{Cluver17}. They have a mean of $\log(\rm{SFR}_{W3}/\rm{M}_{\odot}~\rm{yr}^{-1})=0.72$ and $\log(\rm{SFR}_{W4}/\rm{M}_{\odot}~\rm{yr}^{-1})=0.68$ with standard deviations of 0.30 and 0.41 dex respectively. Panel c) compares the SFRs for each MWA using the W4 and SED-derived SFR indicators. Uncertainties in the SFR relation based on W4 are 0.18 dex and are dominated by scatter in the relation of \citet{Cluver17}, and not from the photometry itself. The mean error in the SED-derived SFRs is 0.11 dex, and these values are shown as a representative error bar in panel c) of Figure~\ref{SFR_plots}. 
There is significant scatter about the 1:1 line, and an offset towards higher W4-derived SFRs for high SED-derived SFRs. Star-shaped points are galaxies that possess known optical AGN or composite regions in their cores, as matched to the MPA-JHU BPT-classified AGN catalogue based on \citet{Brinchmann04}. The AGN in these galaxies may be contributing extra mid-IR flux which is accounted for by the SED-derived SFR indicator, but not in the W4. For this reason, we trust the SED-derived SFRs more than the mid-IR SFR indicator.
For the SED, W3, and W4-derived SFRs, the Milky Way is 1.4$\sigma$, 1.7$\sigma$, and 1.1$\sigma$ away from the means of these distributions, respectively. We list all MWAs in the sample, along with their SFRs, in Table~\ref{data_table1}.

\begin{figure*}
\centering
\begin{subfigure}{0.99\textwidth}
\includegraphics[width=\textwidth]{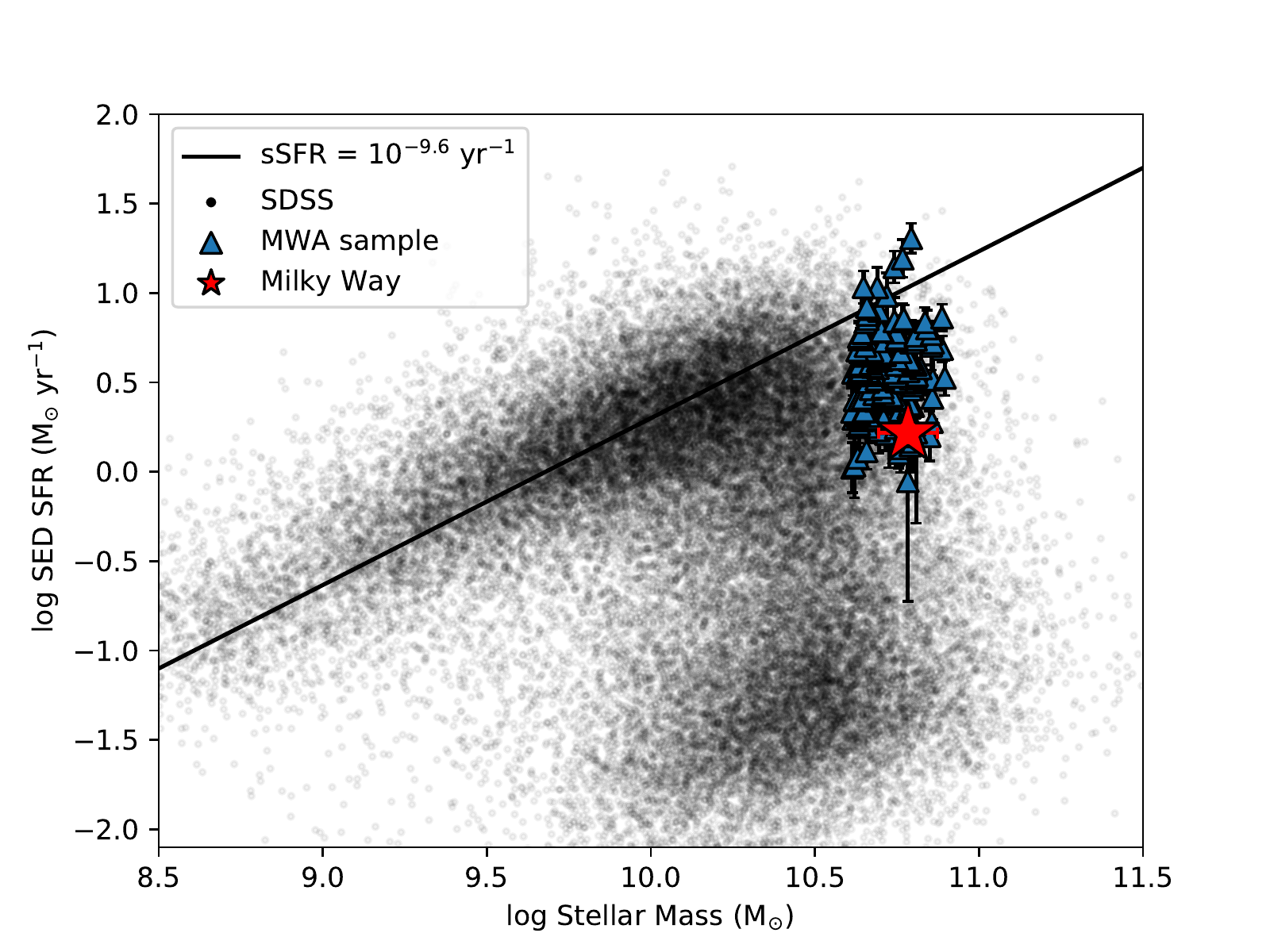} 
\end{subfigure}
\caption{The star formation main sequence. Black points are values from SDSS DR7. Stellar masses are from NSA, and SFRs are the SED-derived values from \citet{Salim18}. MWAs are denoted by blue triangles, and a literature value of the SFR of the Milky Way from \citet{Licquia15} is shown as the red star with error bars from that work. A line of constant $\rm{sSFR} = 10^{-9.6}~\rm{yr}^{-1}$ is used as a proxy for the star formation main sequence line and shown in black. The Milky Way lies slightly below the main sequence, but within 1.4$\sigma$ of the mean of the SED-derived MWA SFRs.} 
\label{SFMS}
\end{figure*}

In Figure~\ref{SFMS}, we plot the SED-derived SFRs of the MWAs along with the rest of SDSS with SED-derived SFRs from \citet{Salim18} and their stellar masses from the NSA, showing the star formation main sequence. As a comparison, the value for the SFR of the Milky Way from \cite{Licquia15} is also plotted as a red star. A line of constant specific star formation rate (sSFR) of $10^{-9.6}~\rm{yr}^{-1}$ is shown in black and may be used as a proxy for the main sequence line. While the tight range in stellar mass is a direct result of the MWA selection technique, we see a large range in derived SFRs for the MWA sample. These galaxies are all high mass spirals, and populate both the `blue cloud' and `green valley' of this parameter space. The Milky Way, while being on the more passive side of this distribution, still comfortably sits within it, with a SFR within 1.4$\sigma$ of the mean SED-derived SFR. 

The advantage of the selection criteria employed in this paper is that they are agnostic to many measurable quantities of the Milky Way. While in this paper we chose to study the star formation rate of MWAs, other quantities such as environment (both local and large-scale), gas fractions, and stellar population distributions are open for study, and we plan to follow these up in a future work. A thorough investigation of the Milky Way in the parameter space of these physical properties will provide a more complete view of the uniqueness of the Milky Way.

In summary, when analogues to the Milky Way are selected solely on non-transient structural properties with an added stellar mass constraint, the mean SFR derived is such that the Milky Way sits well within 2$\sigma$ of the mean of these SFR distributions. For that reason we conclude that given its position just below the main sequence, but relatively small offset from the statistical mean of a sample of MWAs, the Milky Way is not unusual when compared to its immediate peers.

\onecolumn
\begin{longtable}{l c c c c c}
\caption{Table of the Milky Way analogues and their derived SFRs from this work. Here, we present the first ten entries, and the full data table will be available in the online version. When no SED-derived SFR is listed, this value is missing from the \citet{Salim18} catalogue. When no W3 or W4 SFR is listed, $WISE$ photometry wasn't available for this galaxy in the AllWISE source catalogue, usually due to the automated pipeline splitting a single nearby galaxy into two separate sources.}\\ 
\hline
\textbf{Common Name} & \textbf{RA} & \textbf{Dec} & \textbf{$\log(\rm{SFR}_{SED}/$} &  \textbf{$\log(\rm{SFR}_{W3}/$} & \textbf{$\log(\rm{SFR}_{W4}/$}   \\
 & \textbf{(deg)} & \textbf{(deg)} &     \textbf{$\rm{M}_{\odot}~\rm{yr}^{-1})$}  		     &       \textbf{$\rm{M}_{\odot}~\rm{yr}^{-1})$}                     &   \textbf{$\rm{M}_{\odot}~\rm{yr}^{-1}$)}	\\
\hline
  UGC00280               & 7.086   & -0.218  &0.69       &0.62       &0.60      \\
  PGC1142317             & 19.502  & -0.479  &0.76       &0.86       &0.85      \\
  PGC1159961             & 36.135  &0.213   &0.56       &0.72       &0.88      \\
  PGC1132401             & 55.233  & -0.880  &0.36       &0.58       &0.99      \\
  PGC2351360             & 117.193 & 49.720 &0.12       &0.35       &0.10      \\
  PGC2371758             & 117.728 & 50.395 &\--    &0.97       & 1.05     \\
  2MASXJ08012000+1548016 & 120.333 & 15.800 &\--         &\--         &\--        \\
  PGC1998274             & 120.443 & 32.466 &0.68       & 1.16      &0.90      \\
  PGC1322486             & 122.373 & 7.465  &0.56       &0.91       & 1.08     \\
  PGC2253428             & 124.592 & 44.849 & 1.19      & 1.51      & 1.62     \\

\hline
\end{longtable}
\label{data_table1}

\twocolumn

\section{Acknowledgements}

The authors wish to thank Michelle Cluver, Jeffrey Newman, and Steven Bamford for helpful discussions, along with the anonymous referee for providing useful comments that improved the quality of this manuscript.

Funding for the SDSS and SDSS-II has been provided by the Alfred P. Sloan Foundation, the Participating Institutions, the National Science Foundation, the U.S. Department of Energy, the National Aeronautics and Space Administration, the Japanese Monbukagakusho, the Max Planck Society, and the Higher Education Funding Council for England. The SDSS Web Site is http://www.sdss.org/.

The SDSS is managed by the Astrophysical Research Consortium for the Participating Institutions. The Participating Institutions are the American Museum of Natural History, Astrophysical Institute Potsdam, University of Basel, University of Cambridge, Case Western Reserve University, University of Chicago, Drexel University, Fermilab, the Institute for Advanced Study, the Japan Participation Group, Johns Hopkins University, the Joint Institute for Nuclear Astrophysics, the Kavli Institute for Particle Astrophysics and Cosmology, the Korean Scientist Group, the Chinese Academy of Sciences (LAMOST), Los Alamos National Laboratory, the Max-Planck-Institute for Astronomy (MPIA), the Max-Planck-Institute for Astrophysics (MPA), New Mexico State University, Ohio State University, University of Pittsburgh, University of Portsmouth, Princeton University, the United States Naval Observatory, and the University of Washington.

 This publication makes use of data products from the Wide-field Infrared Survey Explorer, which is a joint project of the University of California, Los Angeles, and the Jet Propulsion Laboratory/California Institute of Technology, funded by the National Aeronautics and Space Administration.

    \bibliographystyle{mnras}
  \bibliography{MWAG}
\end{document}